\documentclass[aps,prl,twocolumn,floatfix,showpacs,superscriptaddress]{revtex4}

\usepackage{amssymb}

\usepackage{graphicx}
\usepackage{dcolumn}
\usepackage{bm}
\usepackage{color}
\begin{document}

\title{Magnetic Insulator-Induced Proximity Effects in Graphene: \\Spin Filtering and Exchange Splitting Gaps}
\date{\today}
\author{H.~X.~Yang}
\affiliation{SPINTEC, CEA/CNRS/UJF-Grenoble 1/Grenoble-INP, INAC, 38054 Grenoble, France}

\author{A.~Hallal}
\affiliation{SPINTEC, CEA/CNRS/UJF-Grenoble 1/Grenoble-INP, INAC, 38054 Grenoble, France}

\author{D.~Terrade}
\affiliation{SPINTEC, CEA/CNRS/UJF-Grenoble 1/Grenoble-INP, INAC, 38054 Grenoble, France}

\author{X.~Waintal}
\affiliation{SPSMS-INAC-CEA, 17 rue des Martyrs, 38054 Grenoble France}

\author{S.~Roche}
\affiliation{CIN2 (ICN-CSIC) and Universitat Aut\'{o}noma de Barcelona, Catalan Institute of Nanotechnology, Campus UAB, 08193 Bellaterra, Spain} 
\affiliation{Institucio Catalana de Recerca i Estudis Avan\c cats (ICREA), 08010 Barcelona, Spain}

\author{M.~Chshiev}
\affiliation{SPINTEC, CEA/CNRS/UJF-Grenoble 1/Grenoble-INP, INAC, 38054 Grenoble, France}

\begin{abstract}
We report on first-principles calculations of spin-dependent properties in graphene induced by its interaction with a nearby magnetic insulator (Europium oxide, EuO). The magnetic proximity effect results in spin polarization of graphene $\pi$ orbitals by up to 24 \%, together with large exchange splitting bandgap of about 36 meV. The position of the Dirac cone is further shown to depend strongly on the graphene-EuO interlayer. These findings point towards the possible engineering of spin gating by proximity effect at relatively high temperature, which stands as a hallmark for future all-spin information processing technologies.
\end{abstract}

\maketitle

Heat dissipation has become the bottleneck for further downsizing CMOS devices, and one of the proposed alternative is to switch spin without producing charge currents. In practice, the fabrication of components able to simultaneously inject, manipulate and read out currents based on electron spin stands as an overwhelming material and technological challenge. The combination of semiconductors with magnetic materials remains to date unsuccessful owing to material structural and chemical mismatches~\cite{wolf2001,sarma2004,fert2007, dietl2010}. 

Two-dimensional graphene has demonstrated outstanding physical properties such as exceptional electrical, thermal and mechanical properties~\cite{GrapheneRMPgeim,risegraphene}, but also very long spin diffusion lengths up to room temperature~\cite{Tombros2007,Popinciuc2009, FertAPL,Han2011, Yang2011,Maassen2012,Dlubak2012}. This offers an unprecedented platform for the advent of lateral spintronics in which a complete integration of spin injection, manipulation and detection could lead to ultra-fast electronic circuits compatible with more-than-Moore CMOS and non-volatile low energy MRAM memories \cite{Dery2012}. However, a fundamental challenge lies in the development of external ways to control (gate) the propagation of spin-(polarized) currents at room temperature, in view of designing spin logics devices~\cite{Semenov2007,sarma2011}.

Spin in graphene can be influenced by the presence of local magnetic ordering intentionally generated by material design or defects. For instance edge magnetism has been shown to develop in graphene nanoribbons (a few nanometers wide) for certain edge geometries~\cite{GNRzigzag, refAribbon}, or the hole structure of graphene nanomesh~\cite{GNM} was also theoretically proposed to offer robust and room temperature magnetic states able to affect spin transport~\cite{nanomesh}. A lot of interest is also currently devoted to the tunability of spin-polarized currents and magnetoresistance signals by intentional defects, or depositing atoms or molecules (such as hydrogen~\cite{Soriano2011,McCreary2012} or 3d and 5d metal atoms~\cite{metal,3d1, 3d2,5d,ahe} or large molecules~\cite{refA}). Finally, the growth of graphene on magnetic metallic substrates was also proposed as a route for tailoring graphene spin properties~\cite{GrapheneNi,refAnickle,grapheneNiMagnetism, GrapheneRashba,grapheneDopingWithMetal}. However magnetic conducting substrates, which naturally short circuit the graphene layer, restrict fundamentally the design of novel types of spin switches.

In this Letter, we report tunable magnetic proximity effects induced on graphene by a nearby magnetic insulator. We focus on Europium oxide (EuO) which has been recently successfully grown experimentally on graphene~\cite{EuonGr}. By using $ab$-$initio$  simulations, with both VASP and SIESTA codes, the structure and spin-dependent electronic properties of graphene/EuO junctions are computed. Our findings show that the magnetic interaction induces a large spin polarization of graphene $\pi $-orbitals, leading about 24\% together with a large exchange splitting bandgap of 36 meV.

\begin{figure}
  \includegraphics[width=8 cm]{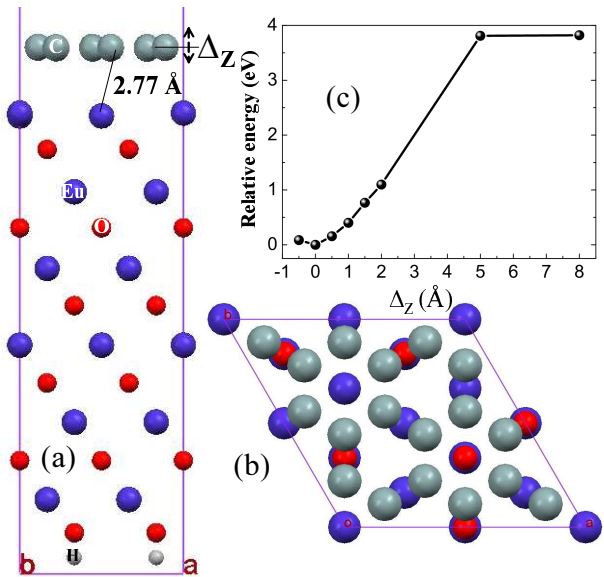}\\
   \caption{(a) side view and (b) top view of the calculated crystalline structures for graphene on top of a six bi-layer 
   EuO film, the bottom of EuO is terminated with hydrogen atoms. (c) relative energy (to the optimized structure) of Graphene/EuO 
   as a function of shifting distance ($\Delta_\mathrm{Z}$) between graphene and substrate.}\label{fig1}
\end{figure}

\begin{figure*}
\includegraphics[width=16 cm]{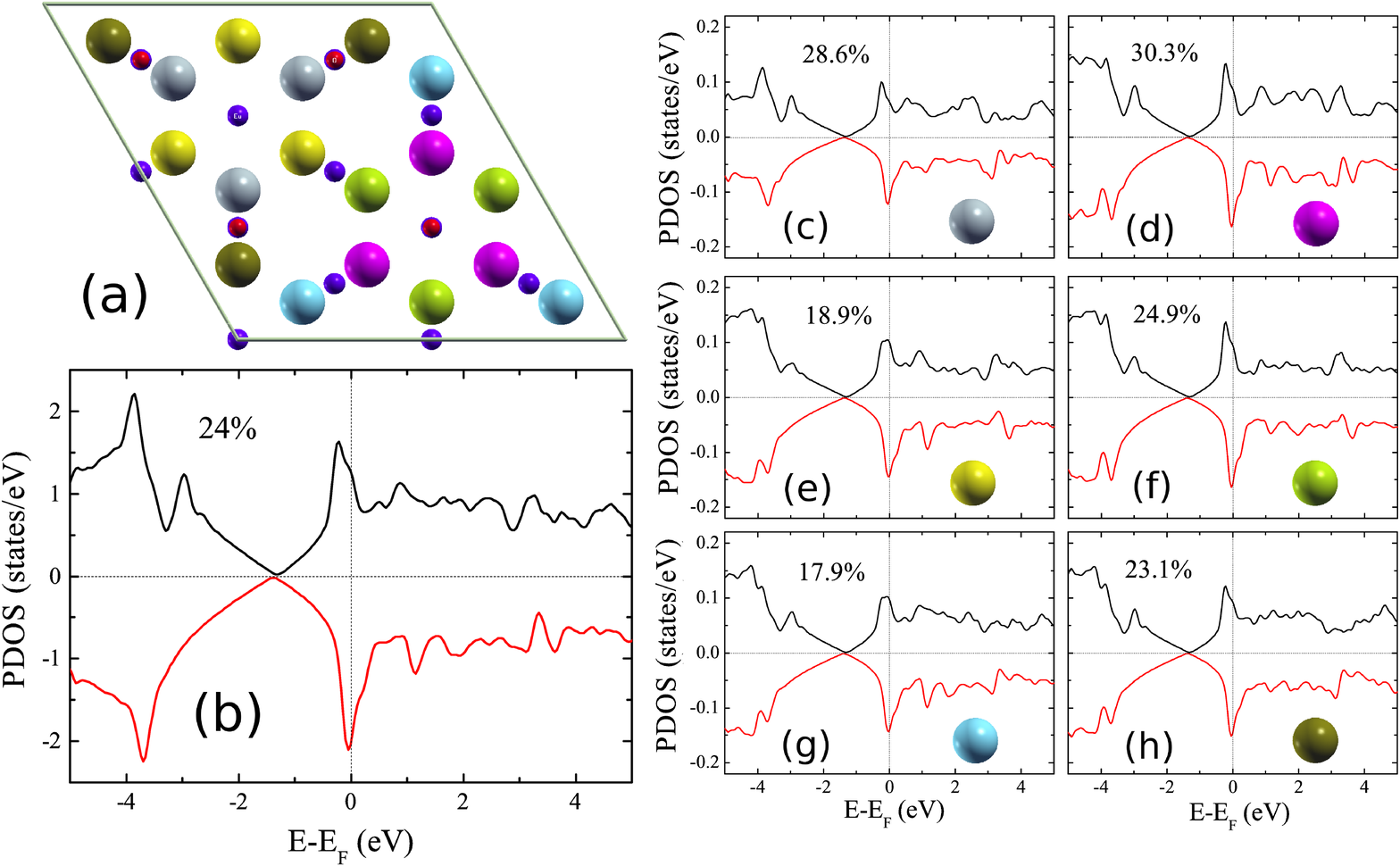}\\
   \caption{(a) The six lattices of graphene on EuO represented with different colors, (b) 
   total density of states of $p_z$ orbital of graphene, (c)-(h) local density of states on each inequivalent atom of the supercell (a). The spin 
   polarizations in (b) to (h) are calculated by comparing the density of states between minority and majority states normalized by the total density of states at Fermi level,
i.e. $p={\frac{n^\downarrow - n^\uparrow}{n^\downarrow + n^\uparrow}}$.}\label{fig2}
\end{figure*}

The Vienna $ab$ $initio$ simulation package (VASP)~\cite{vaspPRB93,vaspPRB96,vaspCMS} is used for structure optimization,
where the electron-core interactions are described by the projector augmented wave method for the pseudopotentials~\cite{pawpotential}, 
and the exchange correlation energy is calculated within the generalized gradient approximation of the Perdew-Burke-Ernzerhof (PBE) 
form~\cite{gga,pbe}. The cutoff energies for the plane wave basis set used to expand the Kohn-Sham orbitals are 520 eV for all 
calculations. A 4 $\times$ 4 $\times$ 1 k-point mesh within Monkhorst-Pack scheme is used for the Brillouin zone integration. 
Structural relaxations and total energy calculations are performed ensuring that the Hellmann-Feynman forces acting on ions 
are less than 10$^{-3}$ eV/\AA. Since Eu is a heavy element with atomic number of 63, and its outer shell (4f$^7$6s$^2$) 
contains 4$f$ electrons, GGA approach fails to describe the strongly correlated localized $4f$ electrons of EuO and predicts a metallic 
ground state of EuO, whereas a clear band gap is observed in experiment ~\cite{EuOfilm,EuObandgapExperiment}. Thus, 
we introduce a Hubbard-U parameter to describe the strong intra-atomic interaction in a screened Hartree-Fock like manner. 
For the parameter choice, we fix on-site Coulomb repulsion and exchange interaction on Eu $4f$ orbital as 8.3 eV and 
0.77 eV, respectively. For oxygen 2$p$ orbitals, the on-site Coulomb and exchange parameters are 4.6 eV and 1.2 eV, 
respectively~\cite{EuOfilm}.

Using GGA+U method, the structure is first optimized from energy considerations and the obtained value of 5.188~\AA~lattice constant is found to be very close to experimental data (5.141~\AA) with an error of only 0.9\% and close to the LDA+U (5.158 \AA). With the optimized lattice, we calculated the density of states for EuO with ferromagnetic state, where a band gap is observed with a value about 1.0 eV. 
This is consistent with the experimental optical absorption gaps of 0.9 and 1.2 eV observed below and above the magnetic transition 
temperature~\cite{EuOgap,Tsymbal}. Here GGA+U method gives better results compared to LDA+U. LDA+U method gives 0.7 eV band gap with 
ferromagnetic spin arrangement, and 1.2 or 1.3 eV for {111} antiferromangetic spin configuration (AFMI) or the NiO-type 
antiferromangetic spin configuration (AFMII), respectively.

We next consider the lattice mismatch between graphene and EuO. If we use experimental values, graphene lattice constant is 2.46 \AA, 
and EuO one is 5.141 \AA. On EuO (111) substrate, a 2 $\times$ 2 
surface unit cell is about 7.2704 \AA, which can fit with a 3 $\times$ 3 unit cell of graphene with lattice mismatch about 1.46\%. If the 
GGA+U optimized lattice constant is used, 5.188 \AA, the mismatch 
is much smaller (less than 1\%). In our calculations, we used the theoretical lattice constant.

With such a reasonable lattice matching, we first evaluated the stability of graphene on EuO surface before studying spintronic properties. Two 
structures with graphene on oxygen terminated surface and Eu terminated surfaces of EuO are considered. For these two configurations, we have the same 
amount of atoms, 18 carbon atoms, 24 oxygen and 24 europium atoms. The calculated total energies are found to be -544.50389 eV and -545.16824 
eV for graphene on O-terminated and Eu-terminated EuO surfaces, respectively. One can see that with Eu-terminated surface, the system is more 
stable with an energy gain of 0.67 eV. Thus, we use the lowest energy configuration of 12 layers of EuO as a substrate. To avoid 
the bottom surface effects on graphene, the bottom oxygen atoms are terminated by hydrogen to simulate graphene on a semi-infinite EuO surface. For 
all calculations, the vacuum length is chosen larger than 20 \AA. The optimized distance between EuO substrate and graphene is shown in 
Figure~\ref{fig1} with a
vertical distance around 2.57 \AA~(nearest C-Eu distance of 2.77 \AA ).

\begin{figure}
  \includegraphics[width=8 cm]{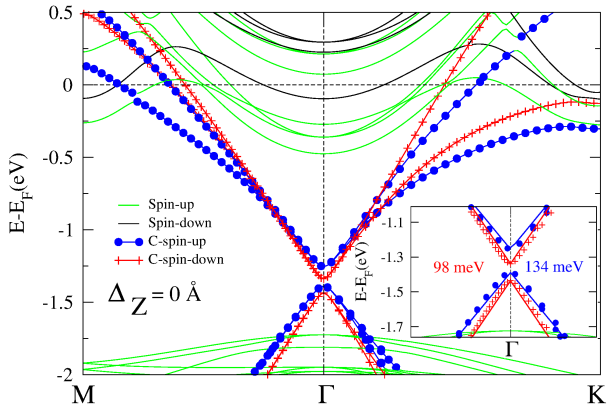}\\
   \caption{Band structure of graphene on EuO. Black and red lines represent spin up and spin down bands, 
   respectively. Inset: zoom around the Dirac cone, the symbols correspond to DFT data while the lines correspond to the fit of Eq.~(\ref{eqgap}).}\label{fig3}
\end{figure}

Using SIESTA package \cite{Soler2002} and the optimized structure of graphene on EuO shown in Figure~\ref{fig1}(a,b), we calculate the local density of 
states for this system [Figure~\ref{fig2}(c-h)] with LDA+U for the exchange correlation functional.
The self-consistent calculations are performed with an energy cutoff of 600 $Ry$ and with a 4 $\times$ 4 $\times$ 1 k-point grid.
A linear combination of numerical atomic orbitals with double-$\zeta$ polarization (DZP) basis set is used.
Due to the existence of EuO substrate, the two sublattices of free 
standing graphene break into six folders as shown in Figure~\ref{fig2}(a) with different colors. In this structure, the calculated magnetic 
moment of surface Eu atoms is found a little bit enhanced, about 7.1 $\mu_B$, compared to the bulk value of 6.9 $\mu_B$. Additionally, the sublayer oxygen 
atoms are found to be spin polarized as well, with magnetic moment of about -0.03 $\mu_B$. The interaction with the magnetic substrate remarkably affects the magnetic properties of graphene. As shown in Figure~\ref{fig2}(b), the average spin 
polarization in the graphene layer is found to be about 24\%. Here, spin polarization is defined as a difference between minority and majority states normalized by the total density of states at Fermi level,
i.e. $p={\frac{n^\downarrow - n^\uparrow}{n^\downarrow + n^\uparrow}}$.
 
Since these 18 carbon atoms are broken into 6 symmetry groups, their contribution to the total spin polarization are different. The spin polarization of magenta atoms can reach up to 30.3\% [Figure~\ref{fig2}(d)], while for the yellow and blue ones, spin polarization is about 18\% [Figure~\ref{fig2}(e,g)]. As shown in Figure~\ref{fig2}(b), the spin polarization is induced on $p_z$ orbital, namely $\pi$ bond.

\begin{figure}
  \includegraphics[width=9 cm]{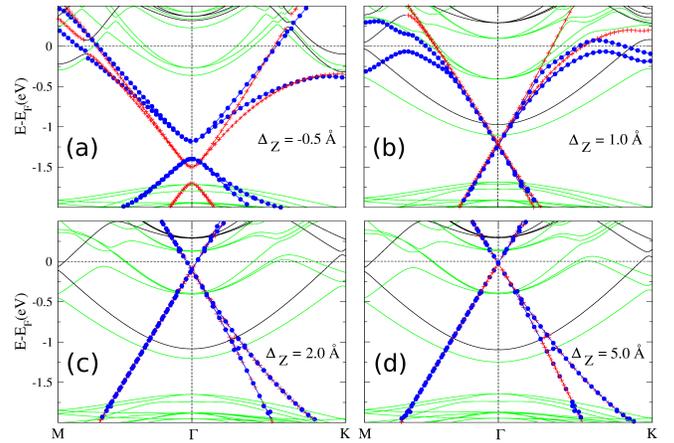}\\
   \caption{Band structures for graphene on EuO with graphene shifted (a) inward (compared to optimized structure) 0.5 \AA, (b) 
   outward 1.0 \AA, (c) outward 2.0 \AA~and (d) outward 5.0 \AA, ~respectively.}\label{fig4}
\end{figure}

We now scrutinize the proximity effect on the graphene band structures. In free standing graphene, the honeycomb structure can be seen as a triangular lattice with a basis of two atoms per unit cell, with 2D lattice vectors ${\mathbf{A_0}=\frac{a_0}{2}(\sqrt3,1)}$ 
and ${\mathbf{B_0}=\frac{a_0}{2}(\sqrt3,-1)}$, 
where $a_0$ is the graphene lattice. Of particular importance for the physics of graphene are the two $\mathbf{K}$ and $\mathbf{K'}$ points at 
the inequivalent corners of the graphene Brillouin zone
${\mathbf{K}=\frac{2\pi}{a_0}(\frac{1}{\sqrt3}, \frac{1}{3})}$ and ${\mathbf{K'}=\frac{2\pi}{a_0}(\frac{1}{\sqrt3}, -\frac{1}{3})}$,
which are called Dirac points. The band dispersion close to the $\mathbf{K}$ (or $\mathbf{K'}$) vector, as $\mathbf{k}=\mathbf{K}+\mathbf{q}$, 
with $|\mathbf{q}|\ll |\mathbf{K}|$, has the form,~\cite{graphitebandtheory}

\begin{equation}\label{equation1}
E_\pm(\mathbf{q})\approx \pm v_F|\mathbf{q}|+O[(q/K)^2]
\end{equation}
where $\mathbf{q}$ is the momentum measured relatively to the Dirac point and $v_F$ is the Fermi velocity, 
given by $v_F=\sqrt{3}ta_0/2$, with a value $v_F\simeq 1\times10^6 m/s$, and $t$ is the nearest-neighbor hopping energy of 2.8 eV.

In the usual case with $\epsilon (\mathbf{q})=q^2/(2m)$, where $m$ is the electron mass, the velocity, $v=k/m=\sqrt{2E/m}$, 
changes with energy. In Equation ~\ref{equation1}, the Fermi velocity does not depend on the energy or momentum~\cite{GrapheneRMPgeim}. 

In case of graphene on magnetic insulator, EuO, as discussed earlier,  the two sublattices have been broken into six  groups [Figure~\ref{fig2}(a)] and there is spin injection to graphene from EuO substrate. The linear dispersion of 
graphene band structure is modified [Figure~\ref{fig3}], with a band gap opening at Dirac point. More interestingly, this degeneracy lifting at the Dirac point is spin dependent: we have fitted the band structure with a simple spin dependent Dirac dispersion relation in presence of a spin dependent mass (gap) term 
\begin{equation}\label{eqgap}
E_\sigma(q)=\pm\sqrt{(\hbar v_\sigma q)^2 + (\Delta_\sigma/2)^2}
\end{equation}
 and obtain gap widths with values of $\Delta_\downarrow=98$ meV and $\Delta_\uparrow=$134 meV for minority and majority states, respectively while the Fermi velocities  $v_\downarrow=1.40\times 10^6 m\cdot s^{-1}$  and $v_\uparrow=1.15 \times 10^6 m\cdot s^{-1}$ are also polarized [see inset of Figure~\ref{fig3}]. The corresponding polarization, around 20\% for both gaps and velocities, is very significant. There is in particular a large energy window (inside the gap region) where the graphene would be 100\% polarized (half-metal) along the majority or minority direction depending on the position of the Fermi level. The observed spin splitting 
is due to the interaction between C-$p_z$ and Eu-4$f$ states. Indeed,  there is a strong peak of polarized Eu-4$f$ state right below 
Fermi level. These polarized states get hybridized with graphene, hence the induced magnetism. This can also be seen from the band structure, where the graphene bands start overlapping with the majority Eu-4$f$ bands. 

Since the interaction between graphene and the substrate is quite weak (as evidenced by the large equilibrium distance of 2.57 \AA~between graphene and EuO layers), it can be further easily affected by the external environment. To mimic the situation of internal pressure and strain, we calculated the electronic properties for varying the interlayer spacing ($\Delta_{\mathrm{Z}}$) as described in Figure~\ref{fig1}(a) and (c). It is found that by interlayer displacement of less than 1 \AA, the total energy of the bilayer system changes only by 0.085 and 0.156 eV/cell for displacement in- and outward of 0.5~\AA, respectively. This weak modification (for supercell contains 18 carbon atoms) corresponds to fluctuations in energy per carbon atom in the order of few meV.

Even though the energetics stability with interlayer spacing $\Delta_{\mathrm{Z}}$ is weakly affected, the impact on the band dispersion of graphene is markedly strong,
as seen in Figure~\ref{fig4}. When compressing the bilayer by 0.5~\AA, more electrons (and spins) are transfered to the graphene layer due to enhanced overlap between C $p_z$ and Eu $4f$ orbitals. Accordingly, the Dirac point is moved deeper inside the valence bands compared to the equilibrium situation (cf. Figs.~\ref{fig3} and~\ref{fig4}(a)). In contrast, for larger layers separation, the Dirac cone is clearly seen to be shifted out from the valence band of EuO, approaching the Fermi level of the system [Fig.~\ref{fig4}(b) and (c)]. Simultaneously, with the shifting of the Dirac point out of EuO valence band, the gap between spin-up and spin-down bands is continuously reduced. Finally, for $\Delta_{\mathrm{Z}}=5$~\AA, the spin up and spin down branches become almost degenerated and the Dirac point crosses the Fermi level, i.e. approaching typical band structure characteristics of isolated graphene~[Fig.~\ref{fig4}(d)]. 

In conclusion,  we have reported first-principles simulations showing that the proximity of a magnetic insulator will induce a strong spin polarization of graphene $\pi$ orbitals. The Europium oxide substrate was found to break the  bipartite lattice of graphene into six inequivalent sublattices, causing variable spin polarizations on the new graphene sublattices with an average spin polarization about 24\%. Simultaneously, a band gap develops at the Dirac point with a large exchange splitting of more than 30 meV, larger than anticipated~\cite{brataas}.
These theoretical findings deserve further experimental demonstration of spin filtering effect and spin-dependent gap in graphene based structures.

$Note$, after the submission of this paper, we just noticed that a successful experimental fabrication of EuO on Graphene \cite{euogr2}.

We thank Profs. Albert Fert, L. Magaud and J. Velev for fruitful discussions. This work was supported by Chair of Excellence Program of the Nanosciences Foundation in Grenoble, France, by French National Research Agency (ANR) Projects NANOSIM-GRAPHENE and NMGEM and by European  Union funded STREP Project CONCEPT-GRAPHENE.


\begin{thebibliography}{200}
\bibitem{wolf2001}
S. A. Wolf, D. D. Awschalom, R. A. Buhrman, J. M. Daughton, S. von Molnar,
M. L. Roukes, A. Y. Chtchelkanova, and D. M. Treger, Science {\bf 294}, 1488 (2001).

\bibitem{sarma2004}
I. $\check{Z}$uti$\acute{c}$, J. Fabian, and S. Das Sarma, Rev. Mod. Phys. {\bf 76}, 323 (2004). 

\bibitem{fert2007}
C. Chappert, A. Fert, and F. Nguyen Van Dau, Nature Mater. {\bf 6}, 813 (2007).

\bibitem{dietl2010}
T. Dietl, Nature Mater. {\bf 9}, 965 (2010).

\bibitem{GrapheneRMPgeim} 
A. H. Castro Neto, F. Guinea, N. M. R. Peres, K. S. Novoselov, and A. K. Geim, Rev. Mod. Phys. {\bf 81}, 109 (2009).

\bibitem{risegraphene} 
A. K. Geim, and K. S. Novoselov, Nature Mater. {\bf 6}, 183 (2007).

\bibitem{Tombros2007}
    N. Tombros,
    C.Jozsa,
    M. Popinciuc,
    H. T. Jonkman, and
    B. J. van Wees
 Nature {\bf 448}, 571 (2007).
 
\bibitem{Popinciuc2009}
M. Popinciuc, C. J$\acute{o}$zsa, P. J. Zomer, N. Tombros, A. Veligura, H. T. Jonkman, and B. J. van Wees, Phys. Rev. B  {\bf 80}, 214427 (2009)


\bibitem{FertAPL} 
B. Dlubak, P. Seneor, A. Anane, C. Barraud, C. Deranlot, D. Deneuve, B. Servet, R. Mattana, 
F. Petroff, and A. Fert, Appl. Phys. Lett. {\bf 97}, 092502 (2010).

\bibitem{Han2011}
W. Han, and R. K. Kawakami, Phys. Rev. Lett.  {\bf 107}, 047207 (2011)

\bibitem{Yang2011}
T.-Y. Yang, J. Balakrishnan, F. Volmer, A. Avsar, M. Jaiswal, J. Samm, S. R. Ali, A. Pachoud, M. Zeng, M. Popinciuc, G. G$\ddot{u}$ntherodt, B. Beschoten, and B. $\ddot{O}$zyilmaz, Phys. Rev. Lett.  {\bf 107}, 047206 (2011)

\bibitem{Maassen2012}
T. Maassen, J. J. van den Berg, N. IJbema, F. Fromm, T. Seyller, R. Yakimova, and B. J. van Wees, Nano Lett. {\bf 12}, 1498 (2012)

\bibitem{Dlubak2012}
B. Dlubak,
M.-B. Martin,
C. Deranlot,
B. Servet, and
S. Xavier, Nature Physics {\bf 8}, 557 (2012).

\bibitem{Dery2012}
H. Dery, H. Wu, B. Ciftcioglu, M. Huang, Y. Song, R. Kawakami, J. Shi, I. Krivorotov, I. Zutic, and L. J. Sham, IEEE Trans. Electron Devices 59, 259 (2012). 

\bibitem{Semenov2007}
Y. G. Semenov, K. W. Kim, and J. M. Zavada, Appl. Phys. Lett. 91, 153105 (2007)

\bibitem{sarma2011}
S. Das. Sarma, S. Adam, E. H. Hwang, and E. Rossi, Rev. Mod. Phys. {\bf 83}, 407 (2011).

\bibitem{GNRzigzag} 
Y.-W. Son, M. L. Cohen, and S. G. Louie, Nature {\bf 444}, 347 (2006).


\bibitem{refAribbon} W. Y. Kim, and K. S. Kim, Nature nanotech. {\bf3}, 408 (2008).

\bibitem{GNM} 
J. Bai, X. Zhong, S. Jiang, Y. Huang, and X. Duan, Nature Nanotech. {\bf5}, 190 (2010).

\bibitem{nanomesh} 
H.-X. Yang, M. Chshiev, D. W. Boukhvalov, X. Waintal, and S. Roche, Phys. Rev. B {\bf 84}, 214404 (2011).

\bibitem{Soriano2011}
D. Soriano, N. Leconte, P Ordejon, J.-Ch. Charlier, J.-J. Palacios, S. Roche, Phys. Rev. Lett. 107, 016602 (2011)

\bibitem{McCreary2012}
K. M. McCreary, A. G. Swartz, W. Han, J. Fabian, and R. K. Kawakami, arXiv:1206.2628.


\bibitem{metal} K. T. Chan, J. B. Neaton, and M. L. Cohen, Phys. Rev. B {\bf77}, 235430 (2008).

\bibitem{3d1} Z. Qiao, S. A. Yang, W. Feng, W.-K. Tse, J. Ding, Y. Yao, J. Wang, and Q. Niu, Phys. Rev. B {\bf82}, 161414(R) (2010).

\bibitem{3d2} J. Ding, Z. Qiao, W. Feng, Y. Yao, and Q. Niu Phys. Rev. B {\bf84}, 195444 (2011).

\bibitem{5d} H. Zhang, C. Lazo, S. Blugel, S. Heinze, and Y. Mokrousov Phys. Rev. Lett. {\bf108}, 056802 (2012). 

\bibitem{ahe} H. Jiang, Z. Qiao, H. Liu, J. Shi, and Q. Niu, Phys. Rev. Lett. {\bf109}, 116803 (2012).

\bibitem{refA} J. Park, S. B. Jo, Y.-J. Yu, Y. Kim, J. W. Yang, W. H. Lee,
H. H. Kim, B. H. Hong, P. Kim, K. Cho, and K. S. Kim, Adv. Mater. {\bf24}, 407 (2012);
W. Y. Kim and K. S. Kim, Acc. Chem. Res. {\bf43}, 111 (2010);
J. W. Yang, G. Lee, J. S. Kim and K. S. Kim, J. Phys. Chem. Lett. {\bf2}, 2577 (2011).

\bibitem{GrapheneNi} 
A. Varykhalov, J. Sanchez-Barriga, A. M. Shikin, C. Biswas, E. Vescovo, A. Rybkin, D. Marchenko, and O. Rader, Phys. Rev. Lett. {\bf 101}, 157601 (2008).

\bibitem{grapheneNiMagnetism}
M. Weser, Y. Rehder, K. Horn, M. Sicot, M. Fonin, A. B. Preobrajenski, E. N. Voloshina, E. Goering, and Yu. S. Dedkov, Appl. Phys. Lett. {\bf 96}, 012504 (2010).

\bibitem{refAnickle} Y. Cho, Y. C. Choi, and K. S. Kim, J. Phys. Chem. C. {\bf115}, 6019 (2011).

\bibitem{GrapheneRashba} 
O. Rader, A. Varykhalov, J. Sanchez-Barriga, D. Marchenko, A. Rybkin, and A. M. Shikin, Phys. Rev. Lett. {\bf 102}, 057602 (2009). 

\bibitem{grapheneDopingWithMetal} 
 G. Giovannetti, P. A. Khomyakov, G. Brocks, V. M. Karpan, J. van den Brink, and P. J. Kelly, Phys. Rev. Lett.{\bf 101}, 026803 (2008).
 
\bibitem{EuonGr} 
D. F. Forster, T. O. Wehling, S. Schumacher, A. Rosch, and T. Michely, New. J. Phys. {\bf 14}, 023022 (2012).

\bibitem{vaspPRB93} 
G. Kresse and J. Hafner, Phys. Rev. B {\bf 47}, 558 (1993).

\bibitem{vaspPRB96} 
G. Kresse and J. Furthmuller, Phys. Rev. B {\bf 54}, 11169 (1996).

\bibitem{vaspCMS} 
G. Kresse and J. Furthmuller, Computational Materials Science {\bf 6}, 15 (1996).

\bibitem{pawpotential} 
P. E. Blochl, Phys. Rev. B {\bf 50}, 17953 (1994).

\bibitem{gga} 
Y. Wang and J. P. Perdew, Phys. Rev. B {\bf 44}, 13298 (1991).

\bibitem{pbe} 
G. Kresse and D. Joubert, Phys. Rev. B {\bf 59}, 1758 (1999).

\bibitem{EuOfilm} 
N. J. C. Ingle and I. S. Elfimov, Phys. Rev. B {\bf 77}, 121202 (2008).

\bibitem{EuObandgapExperiment} 
A. Mauger and C. Godart, Physics Reports {\bf 141}, 51 (1986).

\bibitem{EuOgap} 
J. Schoenes and P. Wachter, Phys. Rev. B {\bf 9}, 309 (1974).

\bibitem{Tsymbal} 
P. V. Lukashev, A. L. Wysocki, J. P. Velev, M. van Schilfgaarde, S. S. Jaswal, K. D. Belashchenko, and E. Y. Tsymbal, Phys. Rev. B {\bf 85}, 224414 (2012).

\bibitem{Soler2002} 
J. M. Soler, E. Artacho, J. D. Gale, A. Garc\`{\i}a, J. Junquera, P. Ordej\'{o}n and Daniel S\'{a}nchez-Portal. J. Phys.-Condes. Matter {\bf 14}, 2745 (2002).

\bibitem{graphitebandtheory} 
P. R. Wallace, Phys. Rev. {\bf 71}, 622 (1947). 

\bibitem{spinGroundStateTheorem}
 E. H. Lieb, Phys. Rev. Lett. {\bf 62}, 1201 (1989).
 
 \bibitem{brataas} 
H. Haugen, D. Huertas-Hernando, and A. Brataas, Phys. Rev. B {\bf 77}, 115406 (2008).

\bibitem{euogr2}
A. G. Swartz, P. M. Odenthal, Y. Hao, R. S. Ruoff, and R. K. Kawakami, ACS Nano, DOI:10.1021/nn303771f (2012)
 
 
\end{thebibliography}
\end{document}